\documentclass[aps,a4paper,prb,eqsecnum,showpacs,floatfix,groupedaddress,twocolumn,merge,elide,flotfix]{revtex4-1}

\usepackage{amsmath}
\usepackage{amsfonts}		
\usepackage[usenames,dvipsnames,svgnames]{pstricks}		
\definecolor{carmine}{rgb}{0.59, 0.0, 0.09}
\definecolor{cardinal}{rgb}{0.35, 0.15, 0.13}

\usepackage{graphicx}		
\usepackage{tikz}			

\usepackage{fourier}
\usepackage[ulem=normalem]{changes}
\setaddedmarkup{\blue{#1}}
\setdeletedmarkup{\red\sout{#1}}

\newcommand{\REM}[1]{}

\usepackage[%
	  unicode=true,%
	  colorlinks=true,%
	  linkcolor=blue,anchorcolor=blue,%
	  breaklinks=true,%
	  citecolor=blue,filecolor=cyan,%
	  menucolor=blue,%
	  breaklinks=true,%
	  urlcolor=blue]{hyperref}
	
\begin{document}
	
\title{
Magnetization steps in the molecular magnet Ni$_4$Mo$_{12}$ revealed by complex exchange bridges}
\author{M.~Georgiev}
\email{mgeorgiev@issp.bas.bg}
\affiliation{Institute of Solid State Physics, Bulgarian Academy of Sciences,
Tsarigradsko Chauss\'ee 72, 1784 Sofia, Bulgaria}
\author{H.~Chamati}
\email{chamati@issp.bas.bg}
\affiliation{Institute of Solid State Physics, Bulgarian Academy of Sciences,
Tsarigradsko Chauss\'ee 72, 1784 Sofia, Bulgaria}

\date{\today}
\begin{abstract} 
We study the behavior of the magnetization and the magnetic susceptibility of molecular 
magnets with complex bridging structure. Our computations are based on
a post-Hartree-Fock method accounting for the intricate network of interatomic bonds 
and an effective spin-like Hamiltonian that captures the essential
magnetic features of magnetic molecules.
The devised method and the constructed Hamiltonian are further employed to characterize the magnetic properties
of the molecular magnet Ni$_4$Mo$_{12}$. The obtained results
reproduce both quantitatively and qualitatively the main features of the 
magnetic spectrum. Furthermore, the computations for the magnetization and the low-field 
susceptibility are in very good agreement with their experimental
counterparts.
In this respect, they improve upon the results obtained
with conventional Heisenberg models.
\end{abstract}
\maketitle

\section{Introduction}\label{sec:introduction}
Molecular magnets are some of the most prominent
examples of physical systems that unveil the quantum origin of magnetism.  
Due to their plain chemical structure and small number
of constituent ions, nanomagnets demonstrate peculiar magnetic and related properties 
that underpin future applications
\cite{hernando_nanomagnetism_1993,gatteschi_molecular_2006,bogani_2008,nasirpouri_nanomagnetism_2011,chamati_theory_2013,liddle_2015} and
challenge scientists working in different fields.

The effects observed in Mn$_{12}$-acetate 
\cite{caneschi_1991,sessoli_1993,jang_2000,regnault_2002,mazurenko_2014,chiesa_2017},
Fe$_8$ based molecular magnets \cite{wernsdorfer_1999,maccagnano_2001,leviant_2014,yaari_2017} and
in the Ni$_4$ clusters
\cite{schnack_observation_2006,kostyuchenko_non-heisenberg_2007,nehrkorn_inelastic_2010,furrer_magnetic_PRB_2010,hubner_2017},
strongly emphasize the unique role of nanomagnets in determining the relation
between electrons' correlations and magnetism.
On the molecular level the existing magnetic exchange processes uniquely 
characterize the nanomagnets' properties \cite{klemm_single-ion_2008,hubner_2017}. 
The number and type of constituing elements,
their bonding and thus the resulting geometric symmetry lies in
the footprints of each nanomagnet, such as
the lightly distorted octahedral magnetic molecules with a central Cr ion 
\cite{salman_2002}, the high spin molecular complex Fe$_{19}$
\cite{blundell_2003}, the molecular wheel Fe$_{18}$ 
\cite{ummethum_2012} with eighteen antiferromagnetically coupled 
spin-$\tfrac52$ ions and the heterometalic Cr$_7$Ni antiferromagnet \cite{timco_2009,wedge_2012,sanna_2017}.
Despite the fact that in addition to the above mentioned
compounds a huge variety of molecular magnets
(see e.g.
\cite{stebler_intra_1982,borras_1999,matsuda_magnetic_2005,procissi_2006,lago_2007,zhao_2009,ghosh_magnetic_2010,minamitani_2016,seikh_2017,zhao_2017}
and references therein)
were synthesized and extensively studied during the last decade,
one could still be fascinated by the richness of their magnetic
properties and the challenges they pose to scientists.

\begin{figure}[b]
	\centering
	\includegraphics[scale=0.97]{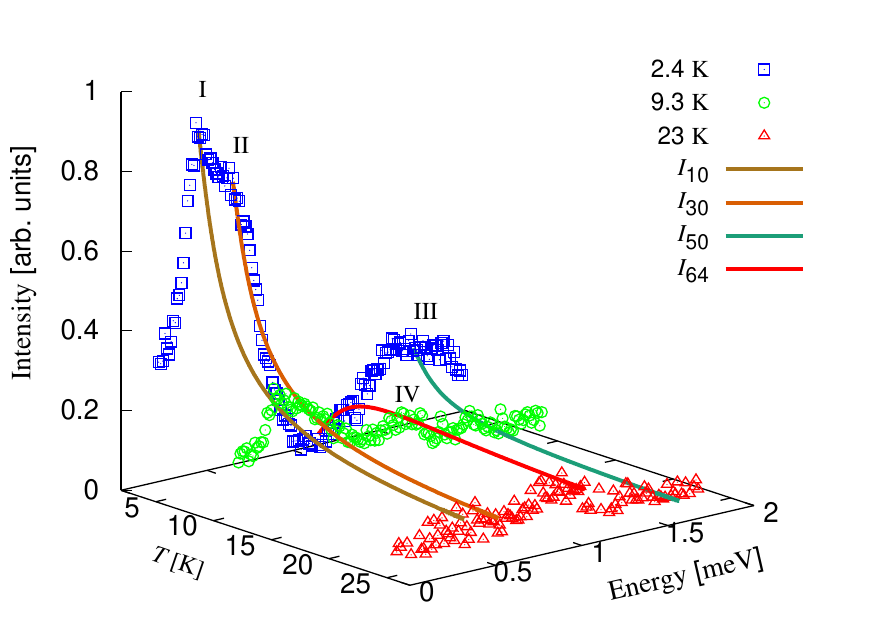}
	\caption{Inelastic neutron scattered intensities for the
	magnetic molecule 
	Ni$_4$Mo$_{12}$ as a function of the temperature and neutrons'
	energy transfer. The experimental data are taken from Ref. 
	\cite{nehrkorn_inelastic_2010}. The curves represent the calculated 
	intensities $I_{n'n}$, with $n$ and $n'$ denoting the number of initial 
	and final states in the transition processes, where $n=0$ is the ground state.
	}
	\label{fig:NiExp3d} 
\end{figure}
The intricate magnetic features of the molecule
[Mo$_{12}$O$_{30}$($\mu_2$-OH)$_{10}$H$_2$(Ni(H$_2$O)$_3$)$_4$]
(Ni$_4$Mo$_{12}$), with four Ni magnetic centers occupying the 
vertices of a slightly distorted tetrahedron \cite{muller_2000},
has been the subject of many investigations
\cite{nehrkorn_inelastic_2010,kostyuchenko_non-heisenberg_2007,furrer_magnetic_PRB_2010,hubner_2017}.
In Ref. \cite{nehrkorn_inelastic_2010}, a very general
Heisenberg model involving different sorts of magnetic interactions
among localized spins and the Hubbard model with localized electrons
were used to analyzed INS data.
Both studied models were proven unable to appropriately explain the
magnetism in the named molecular magnets.
Further analysis of the magnetic spectra of Ni$_4$Mo$_{12}$,
formulated as 
[Mo$_{12}$O$_{28}$($\mu_2$-OH)$_{9}$($\mu_3$-OH)$_{3}$(Ni(H$_2$O)$_3$)$_4$],
was performed in Ref. \cite{furrer_magnetic_PRB_2010}, where in terms of the Heisenberg model a
``naive'' spin coupling scheme with two arbitrary coupling parameters
of the isotropic Heisenberg model was used.
The physical reason lying behind this parametrization was not
discussed. A shortcoming of
the proposed model is that it provides a single intensity curve for all three
low-temperature magnetic peaks. Moreover, even after including a
single ion anisotropy with yet another running parameter, the magnetic
features of Ni$_4$Mo$_{12}$ could not be fully determined.
A more recent study \cite{hubner_2017} showed that electron
correlations are the driving mechanisms behind magnetism of
four-center transition-metal clusters.
Thus, despite the efforts
an overall theoretical description and deep understanding of the
magnetic spectrum obtained via 
inelastic neutron scattering (INS) experiments is still an open question. 

In an attempt to elucidate the mechanisms underpinning the
magnetic properties of this molecule we revisited the reported 
experimental data and proposed an alternative
approach
\cite{georgiev_mexchange_2019,georgiev_trimer_2019,georgiev_epjb_2019}
that helped explaining 
the details of the magnetic spectrum obtained via INS, see Fig. \ref{fig:NiExp3d}.  
Within our approach the four magnetic excitations, I, II, III and IV, with energies
$\Delta_{\mathrm{I}}=0.4$, $\Delta_{\mathrm{II}}=0.6$, 
$\Delta_{\mathrm{III}}=1.7$ and $\Delta_{\mathrm{IV}}=1.15$ meV, and 
INS intensities $I_{10}$, $I_{30}$, $I_{50}$ and $I_{64}$, respectively, 
are uniquely characterized. All features in the magnetic spectrum,
such as splitting 
and broadened, were revealed. For further details the reader may 
consult Ref. \cite{georgiev_epjb_2019}.

In order to interpret the experimental measurements of the
magnetization and the susceptibility obtained for Ni$_4$Mo$_{12}$
reported in Ref. \cite{schnack_observation_2006},
the authors relied on electron paramagnetic 
resonance measurements and explored the relevant magneto-optical 
properties.
They combined the isotropic Heisenberg Hamiltonian with a biquadratic 
and a single-ion anisotropy terms in addition to field dependent coupling parameters.
This allowed them to qualitatively reproduce the magnetization and susceptibility 
measurements data but failed to explain the inelastic neutron
scattering experiments \cite{nehrkorn_inelastic_2010}. 
Another study of the magnetization and the susceptibility
\cite{kostyuchenko_non-heisenberg_2007} proposed a non-negligible contribution of 
the three-body spin interaction term.
To reproduce the behavior of the field dependent magnetization and the associated
susceptibility of the compound under consideration, it was suggested
\cite{furrer_magnetic_PRB_2010} to account for single-ion anisotropy,
as well.
However, no firm physical grounds were reported in support of
the pure technical procedure leading to the parametrization
of Ref. \cite{furrer_magnetic_PRB_2010}. Furthermore it does not
account for the background emanating from delocalized electrons.


In this article we report theoretical results on the magnetization and
the magnetic susceptibility results
for the tetramer Ni$_4$Mo$_{12}$ based on the
proposed in \cite{georgiev_epjb_2019} spin-like Hamiltonian taking
into account the Zeeman term. 
From the physical point of view the present model opens a new
perspective on the theoretical studies of the magnetic spectrum, magnetization and magnetic
susceptibility of Ni$_4$Mo$_{12}$. Notice that in contrast to the
introduced in Ref.
\cite{furrer_magnetic_PRB_2010} picture of strongly localized electrons, 
the model considered in this paper views the electrons as delocalized and hence assures the role of the 
bridging structure in the exchange processes. This is in concert with
other investigations on clusters with four Ni centers
\cite{kostyuchenko_non-heisenberg_2007,nehrkorn_inelastic_2010,hubner_2017,georgiev_epjb_2019}.
We would like to anticipate that
the obtained results are in good qualitative and fairly quantitative
agreement with the corresponding experimental
measurements data \cite{schnack_observation_2006}.
The analysis suggests that the observed shifting in magnetization steps 
is associated to variations in the correlations between delocalized electrons due to the
presence of externally applied magnetic field.
The influence of the field is indirect and differ by a contribution
emanating from the Zeeman term. 
It originates from the interaction between the molecules intrinsic magnetic 
vector potential and the magnetic vector potential of the external
field,
and it is especially underlined due to the electrons' delocalization.
The resulting interaction terms contribute to the correlation functions derived from a 
post-Hartree-Fock method and thus enters into the spin-like Hamiltonian.
Moreover, the approach proposed here leads to improvements
upon that based on the conventional isotropic Heisenberg model.
For the sake of comparison we computed the magnetization
and magnetic susceptibility with the aid of the isotropic Heisenberg
model, as well as the Hamiltonian involving anisotropic in
space spin-spin interaction and an axial single-ion anisotropy
proposed in Ref. \cite{furrer_magnetic_PRB_2010}.

The rest of this paper is organized as follows: In Section
\ref{sec:spinmodel} we write down the Hamiltonian and 
the model parameters relevant to the present study.
In Sections \ref{sec:nickel} we
obtain the explicit expression of the energy spectrum as a function
of all running parameters used to study the magnetization and the magnetic susceptibility of the 
Ni$_4$Mo$_{12}$. In Section \ref{sec:magnetization}
we determine the values of the model parameters fitted to the experimental results. 
A summary of the results obtained throughout this paper is
presented in Section \ref{sec:conclusion}.

\section{The effective model}

\subsection{The spin-like Hamiltonian} \label{sec:spinmodel}

In order to investigate the magnetic properties of the
tetramer we apply the formalism presented in Ref. \cite{georgiev_epjb_2019}
by accounting for the action of the externally applied magnetic field.
Thus, we consider the following spin-like Hamiltonian
\begin{equation}\label{eq:AddHamiltonian}
\hat{\mathcal{H}} = 
\sum\limits_{i \ne j}^{} 
J_{ij} \hat{\boldsymbol{\sigma}}_i\cdot\hat{\mathbf{s}}_j-
\mu_{\mathrm{B}} B\sum_{i}g^\alpha_i\hat{\mathrm{s}}^\alpha_i,
\end{equation} 
where $J_{ij}=J_{ji}$ are effective constants, the operator
$
\hat{\boldsymbol{\sigma}}_i =
(\hat{\sigma}^x_i, \hat{\sigma}^y_i, \hat{\sigma}^z_i)
$ 
indirectly accounts for the differences in electrons' distribution with
respect to the 
$i$-th magnetic center with spin operator $\hat{\mathbf{s}}_i$, $\mu_{\mathrm{B}}$ is the Bohr magneton, $g^\alpha_i$ 
is the $\alpha$ component of the corresponding \textit{effective} $g$-factor and $B$ is the 
magnitude of the external magnetic field 
oriented along a preselected magnetic easy axis $\alpha\in\{x,y,z\}$.
Moreover, for each total spin multiplet
the $\sigma$ operators will also account for the rate at which the electrons' correlations
alter due to the presence of an externally applied magnetic field.

We would like to point out that the effective $g$-factor differs from the $g$-tensor known in the
theory of magnetism.
Thus, in contrast to the widely applied $g$-tensor, derived via the quantum perturbation theory, 
the effective $g$-factor resulting from the post-Hartree-Fock method is a three component vector
$\mathbf{g}_i=(g^x_i,g^y_i,g^z_i)$. The reason is that we use a variational method and
the electrons are regarded as delocalized to a large extent, occupying molecular
orbitals.
The technical details behind the derivation of $\mathbf{g}$ vector 
lies bend the objective of the present article and will be a subject 
of separate paper \footnote[1]{M. Georgiev and H. Chamati, under
preparation}.
However, following the short review in Ref. \cite{georgiev_epjb_2019} explaining 
the main steps leading to the construction of the spin-sigma bilinear form in 
\eqref{eq:AddHamiltonian} we can give an explicit representation of
the effective $g$-factor.

It is worth mentioning that the applied formalism is based on a 
multi-configuration self consistent field method 
\cite{roos_complete_2007,szalay_multicon_2012,vonci_magnetic_2017}
that in turn relies on the molecular orbital theory 
\cite{fleming_molecular_2009,albright_orbital_2013} 
as the main approach in describing the interatomic bonding.
The initial canonical Hamiltonian that leads to \eqref{eq:AddHamiltonian}
takes into account the electrons' kinetic energy, the 
electron-electron and electron-nuclei interactions in addition to the influence of
externally applied magnetic field on the electrons' motion.
In addition to the action of the external magnetic field, 
each electron is viewed as interacting with an intrinsic local 
magnetic field that originates from the orbital and spin angular momenta of
all remaining electrons. 
All electrons in the system are considered as delocalized
in terms of the molecular orbital theory. Thus, they occupy molecular orbitals 
$\phi_{n,m_i}(\mathbf{r}_i)$, with $n\in\mathbb{N}$, represented as a
linear combinations of atomic orbitals 
$\psi^{\eta}_{\mu_{\eta i},m_i}(\mathbf{r}_i)$, where $\mathbf{r}_i$ are the 
coordinates of the $i$-th electron, $\mu_{\eta i}$ denotes the electronic shell 
and subshell with respect to the $\eta$ nucleus and $i$-th electron, 
$m_i$ is the spin magnetic quantum number of the $i$-th electron.
In the considered method, different exchange bridges may favor different  
electrons' distributions and thus configurations.
Therefore, the corresponding state functions are given by a linear combination of
Slater determinants with elements $\phi_{n,m_i}(\mathbf{r}_i)$ and are 
symmetrized in accordance to the spin quantum numbers $s_{ij}$ 
of all electron pairs and with respect to all 
probable electrons' distributions along all exchange bridges.


Within the assumption
of delocalized electrons the effective $g$-factor 
may be derived from the interaction between the magnetic vector potential of the 
externally applied field and the magnetic vector potentials associated to the 
orbital and spin magnetic moments of the constituent electrons. 
These interactions result from the relevant electrons' generalized momenta.
In terms of $N$ electrons with $N-2$ pairs, the effective $g$-factor 
associated with the 
$(N-1)$-th and $N$-th unpaired by orbitals electrons is given by
\begin{equation*}
	g_\alpha=\frac{1}{2}\sum_{\tau}\big|c_\tau\big|^2
	\left(g^{\alpha,\tau}_{N-1}+g^{\alpha,\tau}_{N}\right),
\end{equation*}
where the coefficient $c_\tau$ accounts for the probability of observing 
both electrons in one of the three possible triplet configurations 
associated with a set of molecular orbitals 
$\tau=(n,\ldots,n')$, with $n$ denoting the orbital's number. 
Further, for all $\alpha\in\{x,y,z\}$ we have 
\begin{equation*}
	g^{\alpha,\tau}_{N}=
	g_en_\alpha-
	g_e\frac{e^2\mu_0}{8\pi^2m_e}\sum^{N-1}_{j=1}
	\bigg\langle 
	\frac{\alpha_{j}\left(\mathbf{n}\cdot\mathbf{r}_{Nj}\right)-
		n_\alpha\left(\mathbf{r}_{j}\cdot\mathbf{r}_{Nj}\right)}{r^{3}_{Nj}}
	\bigg\rangle_\tau,
\end{equation*}
where $g_e$ is the electron's spin $g$-factor, $e$ is the elementary charge,
$\mu_0$ is the magnetic permeability,
$m_e$ is the electron's rest mass, $\mathbf{n}=(n_x,n_y,n_z)$ 
is a unit vector such that $B=\mathbf{n}\cdot\mathbf{B}$,
$\mathbf{r}_{j}=(x_j,y_j,z_j)$ are the coordinates of the $j$-th electron
and $r_{Nj}$ is the distance separating the $N$-th and the $j$-th electrons.

\subsection{Properties of the $\sigma$ operators}\label{sec:sigma}

For the sake of completeness, we present the properties of $\sigma$ operators relevant 
to the present study. Additional details can be found
in Refs. \cite{georgiev_mexchange_2019,georgiev_epjb_2019}.

The components of $\sigma$ operator are such that 
for all $i$ and $\alpha \in \{x,y,z\}$, we have
\begin{equation}\label{eq:Sigma_i}
\hat{\sigma}^\alpha_i \lvert \ldots ,s_i,m_i, \ldots \rangle
	=
a^{s_i,m_i}_i \hat{\mathrm{s}}^{\alpha_{\phantom{i}}}_i 
\lvert \ldots ,s_i,m_i, \ldots \rangle,
\end{equation}
where $a^{s_i,m_i}_i \in \mathbb{R}$.

When the $i$-th and $j$-th spin centers are coupled, with total 
spin operator $\hat{\mathbf{s}}_{ij}=\hat{\mathbf{s}}_{i}+\hat{\mathbf{s}}_{j}$,
one has the corresponding total $\sigma$-operator $\hat{\boldsymbol{\sigma}}_{ij}$.
Hence, similar to Eqs. \eqref{eq:Sigma_i},
for all $i\ne j$ and $\alpha\in\{x,y,z\}$, we have
\begin{equation}\label{eq:Sigma_ij}
\hat{\sigma}^\alpha_{ij} \lvert\ldots ,s_{ij},s,m\rangle
=
a^{s,s_{ij},m_{ij}}_{ij} \hat{\mathrm{s}}^{\alpha_{\phantom{j}}}_{ij} 
\lvert\ldots ,s_{ij},s,m\rangle,
\end{equation}
where $a^{s,s_{ij},m_{ij}}_{ij} \in \mathbb{R}$. 
The individual $\sigma$-operators from the $(i,j)$-th spin pair share 
the same coefficient. Thus, for any $i\ne j$ and $\alpha\in\{x,y,z\}$ one gets
\begin{equation}\label{eq:Sigma_k}
\hat{\sigma}^{\alpha}_i\lvert\ldots ,s_{ij},s,m\rangle
	=
a^{s,s_{ij},m_{ij}}_{ij} \hat{\mathrm{s}}^{\alpha_{\phantom{j}}}_i 
\lvert\ldots ,s_{ij},s,m\rangle.
\end{equation}

In the presence of an external magnetic field the constraints in Ref.
\cite{georgiev_epjb_2019}, obtained without magnetic field need to be modified.
Since according to the underlying post-Hartree-Fock method the
electrons'
correlations are field dependent, 
in the considered case these constraints enter a more general form.
The equations relating $\sigma$ and spin operators read
\begin{subequations}\label{eq:Constraint_ij}
\begin{equation}\label{eq:SigmaZ_ij}
\hat{\sigma}^z_{ij} \lvert\ldots,s_{ij},s,m \rangle =
h^s_{ij} \, \hat{\mathrm{s}}^z_{ij} \lvert\ldots,s_{ij},s,m \rangle,
\end{equation}
\begin{equation}\label{eq:SigmaSquare_ij}
\hat{\boldsymbol{\sigma}}^2_{ij} \lvert\ldots,s_{ij},s,m \rangle =
\left( h^s_{ij} \right)^2 s_{ij}(s_{ij}+1) 
\lvert\ldots,s_{ij},s,m \rangle,
\end{equation}
\end{subequations}
where the parameter $h^s_{ij}\in\mathbb{R}$ accounts for the changes in electrons' 
correlations due to the indirect action of the externally applied magnetic field.
Such an influence alter the energy spectrum obtained for $B=0$ and differ from the 
contribution associated to the Zeemen term.
Notice that in the absence of external magnetic field, for all $s$ and $s_{ij}$ one has 
$h^s_{ij}=1$. 

Using Eqs. \eqref{eq:Sigma_ij} and the constraints \eqref{eq:Constraint_ij} 
we distinguish four cases:

\textbf{(1)} For $s_{ij}\ne 0$ and $m_{ij}\ne 0$, or when the value of $m_{ij}$ 
cannot be determined, we have
$
a^{s,s_{ij},m_{ij}}_{ij}=h^s_{ij}.
$

\textbf{(2)} When $s_{ij} \ne 0$ and $m_{ij}=0$, then from Eq. 
\eqref{eq:SigmaSquare_ij}
it follows that
\begin{equation}\label{eq:as0_ij}
a^{s,s_{ij},0}_{ij}=\pm h^s_{ij}.
\end{equation}

\textbf{(3)} In the presence of a singlet bond, $s_{ij} = 0$, one has 
unconstrained parameters such that for the set of coefficients 
$c^{n}_{ij}\in\mathbb{R}$,
\begin{equation}\label{eq:a00_ij}
a^{s,0,0}_{ij}\in\big\{h^s_{ij} c^{n}_{ij}\big\}_{n\in\mathbb{N}}.
\end{equation} 

\textbf{(4)} In the case of a total singlet $s=0$ and $s_{ij} = 0$
the system resemble a closed shell system with unique electron configuration.
As a result, the effective parameter will have a unique value 
\begin{equation}\label{eq:a000_ij}
a^{0,0,0}_{ij}=h^0_{ij}.
\end{equation}

For more details on the quantities $c^{n}_{ij}$ see Ref.
\cite{georgiev_epjb_2019}.
In general, within Hamiltonian \eqref{eq:AddHamiltonian} one 
relies on two classes of parameters. The spectroscopic parameters $c^n_{ij}$ 
that can be fitted according to performed spectroscopic measurements and the 
parameters $h^s_{ij}$ that account for the effect of externally applied magnetic filed.
The latter can be fixed with respect to the magnetization and magnetic 
susceptibility measurements.

\section{N\lowercase{i}$_4$M\lowercase{o}$_{12}$} \label{sec:nickel}
\subsection{The Hamiltonian}

The spin-sigma bilinear form in \eqref{eq:AddHamiltonian} is derived
in a such a way to be valid for arbitrary magnetic field. 
Therefore, for calculating the energy spectrum for $B\ne0$,
we take into account the applied spin 
coupling scheme discussed in Ref. \cite{georgiev_epjb_2019}.
In other words we have Ni1-Ni2 and Ni3-Ni4 spin bonds and
the eigenstates $\lvert s_{12},s_{34},s,m \rangle$. Since
the four Ni spin centers are indistinguishable, for $i,j=1,\ldots,4$
we set $J_{ij}=J$. 
Furthermore, we distinguish two $\sigma$ bond operators, $\hat{\boldsymbol{\sigma}}_{12}$
and $\hat{\boldsymbol{\sigma}}_{34}$, corresponding to Ni1-Ni2 and Ni3-Ni4 spin 
pairs, respectively.
According to \eqref{eq:Sigma_ij} and
\eqref{eq:Sigma_k} one has the spin bond
parameters $a^{s,s_{12},m_{12}}_{12}$ and $a^{s,s_{34},m_{34}}_{34}$ associated 
with Ni1-Ni2 and Ni3-Ni4 spin pairs, respectively. 
In our previous work \cite{georgiev_epjb_2019}, with the aid of the spectroscopic 
parameters $c_{12}$ and $c_{34}$, see also \eqref{eq:a00_ij}, 
we were able to unveil the driving exchange mechanism behind the magnetic spectrum
of Ni$_4$Mo$_{12}$. 
Now, taking in to account the field parameters $h^s_{12}$ and $h^s_{34}$ we can
shed more light on the way the external magnetic field affect the electrons'
correlations in this magnetic molecule.

Using \eqref{eq:AddHamiltonian} and selecting the $z$-axis as the magnetic 
easy axis we end up with the Hamiltonian
\begin{align}\label{eq:NickelHamilton}
\hat{\mathcal{H}} = \  & J 
\left( 
\hat{\boldsymbol{\sigma}}_1\cdot\hat{\mathbf{s}}_2 + 
\hat{\boldsymbol{\sigma}}_2\cdot\hat{\mathbf{s}}_1 +
\hat{\boldsymbol{\sigma}}_3\cdot\hat{\mathbf{s}}_4 +
\hat{\boldsymbol{\sigma}}_4\cdot\hat{\mathbf{s}}_3
\right) 
\nonumber \\ 
& + J 
\left( 
\hat{\boldsymbol{\sigma}}_{12}\cdot\hat{\mathbf{s}}_{34} +
\hat{\boldsymbol{\sigma}}_{34}\cdot\hat{\mathbf{s}}_{12}
\right) - g\mu_{\mathrm{B}}B \, \hat{\mathrm{s}}^z,
\end{align}
where $\hat{\mathrm{s}}^z$ is the total spin $z$ component and since all
Ni centers are indistinguishable for $i=1\ldots,4$ we may set $g=g^z_i$. 

With the aid of Hamiltonian \eqref{eq:NickelHamilton} the
total spin $s$ remains a good quantum number and hence the magnetic
molecule is fully described by a total of eighty one eigenstates.

\begin{figure*}[!ht]
\centering
\includegraphics[scale=1]{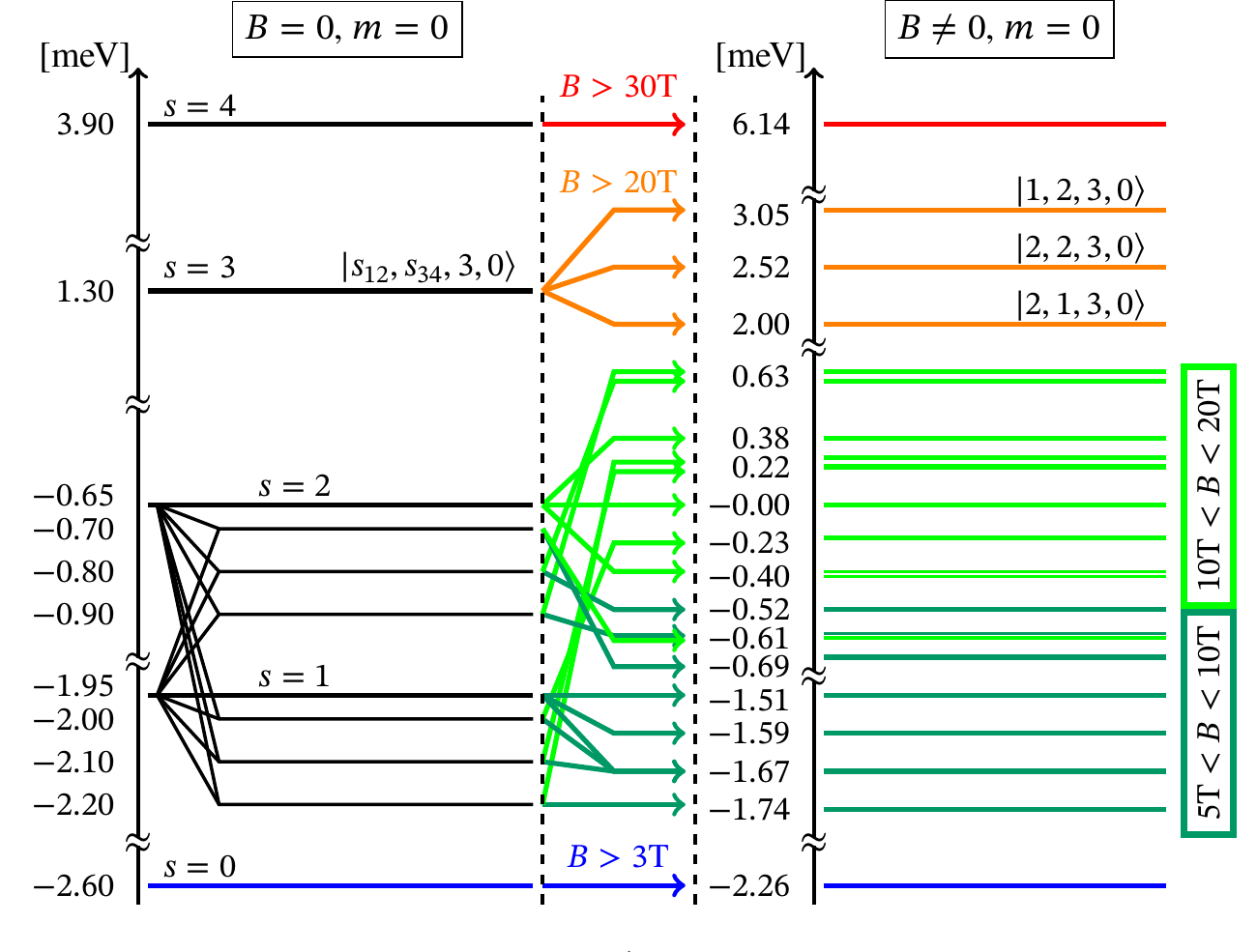}
\caption{
Comparison between the energy spectra of Ni$_4$Mo$_{12}$ for
$B=0$ on the left and $B\ne0$ on the right assuming
only the nonmagnetic states, $m=0$. The blue and red
colors mark the
ground and higher energy levels, respectively. The dark green lines
depict the triplet, the green one stands for the quintet energy levels.
Orange lines correspond to the septet energy levels. A detailed
representation of the initial spectrum, placed on the left, is given
in Ref. \cite{georgiev_epjb_2019}. The domain between both dashed lines
shows the splitting paths of the initial spectrum. The right hand side
spectrum is obtained according to the magnetization and susceptibility
measurements reported in Ref. \cite{schnack_observation_2006}.}
\label{fig:EnSpectrum}
\end{figure*}

\subsection{Energy spectrum}\label{sec:energy}
Bearing in mind that the model parameters 
$a^{s,s_{12},m_{12}}_{12}$ and $a^{s,s_{34},m_{34}}_{34}$ 
take discrete values, the energy spectrum of Hamiltonian 
\eqref{eq:NickelHamilton} can be generalized by the expression
\begin{align}\label{eq:NickelSSEigenstatesB}
	E^m_{s_{12},s_{34},s} = & \
Ja^{s,s_{12},m_{12}}_{12}
\big[ 
s_{12}(s_{12}+1)-2s_0(s_0+1)
\big]
\nonumber \\
 &+
Ja^{s,s_{34},m_{34}}_{34}
\big[ 
s_{34}(s_{34}+1)-2s_0(s_0+1)
\big]
\nonumber \\
	& +
\tfrac{1}{2}J\left( 
a^{s,s_{12},m_{12}}_{12}+a^{s,s_{34},m_{34}}_{34}
\right)\big[ s(s+1)-s_{12}(s_{12}+1)
\nonumber \\
 & -s_{34}(s_{34}+1)
\big]
- g\mu_{\mathrm{B}}Bm.
\end{align}
The eigenvalues in \eqref{eq:NickelSSEigenstatesB} are not explicit functions of
the scalars in \eqref{eq:a00_ij}. 
However, for convenience their dependence on the number $n$
will be explicitly underlined only in the case when $n>1$.

\begin{table}[!t]
	\caption{
		The values of the model parameters used to characterize the magnetic properties of Ni$_4$Mo$_{12}$.
		The evaluation of the spectroscopic parameters $J$, $c^1_{12}$, 
		$c^2_{12}$ and $c_{34}$ is achieved within the neutron spectroscopy
		data of Ref. \cite{georgiev_epjb_2019}. All ``$h$'' parameters
		are fitted in accordance to the experimental data from the
		magnetization measurements depicted on Fig. \ref{fig:MB}.
	}
	\label{tab:Param}
	\begin{center}
		\begin{tabular}{lcccccc}
			\hline\hline\noalign{\vspace{1pt}}
			$B$ [T] & $0$ & $0-4.5$ & $4.5-8.9$ & $8.9-20.1$ & $20.1-32$ & $>32$ \\
			\noalign{\vspace{1pt}}\hline\noalign{\vspace{1pt}}
			$M/g\mu_{\mathrm{B}}$ & $0$ & $0$ & $1$ & $2$ & $3$ & $4$ \\
			\noalign{\vspace{1pt}}\hline\noalign{\vspace{1pt}}
			$J$~[meV] & $0.325$ &  &  &  & & \\
			\noalign{\vspace{1pt}}
			$c^1_{12}$ & $1.192$ &  &  &  & & \\
			\noalign{\vspace{1pt}}
			$c^2_{12}$ & $1.115$ &  &  &  & & \\
			\noalign{\vspace{1pt}}
			$c_{34}$ & $1.038$ &  &  &  & & \\
			\noalign{\vspace{1pt}}
			$h^0_{12}$ & $1$ & $0.945$ &  &  & &\\
			\noalign{\vspace{1pt}}
			$h^0_{34}$ & $1$ & $0.793$ &  &  & & \\
			\noalign{\vspace{1pt}}
			$h^1_{12}$ & $1$ &  & $0.758$ &  & & \\ 
			\noalign{\vspace{1pt}}
			$h^1_{34}$ & $1$ &  & $0.879$ & & & \\
			\noalign{\vspace{1pt}}
			$h^2_{12}$ & $1$ &  &  & $-0.289$ & & \\
			\noalign{\vspace{1pt}}
			$h^2_{34}$ & $1$ &  &  & $\phantom{-}0.315$ & & \\
			\noalign{\vspace{1pt}}
			$h^3_{12}$ & $1$ &  &  &  & $1.532$ & \\
			\noalign{\vspace{1pt}}
			$h^3_{34}$ & $1$ &  &  &  & $2.347$ & \\
			\noalign{\vspace{1pt}}
			$h^4_{12}$ & $1$ &  &  &  &  & $1.554$ \\
			\noalign{\vspace{1pt}}
			$h^4_{34}$ & $1$ &  &  &  &  & $1.595$ \\
			\noalign{\smallskip}\hline\hline
		\end{tabular}
	\end{center}
\end{table}

According to Ref. \cite{georgiev_epjb_2019} and the analysis based on the inelastic neutron 
scattering experiments reported in 
\cite{nehrkorn_inelastic_2010,furrer_magnetic_PRB_2010} 
the ground state of this nanomagnet appears to be the singlet 
$\lvert 1,1,0,0 \rangle$.
Taking into account \eqref{eq:Constraint_ij} for the ground state
energy, we obtain
\[
	E^{0}_{1,1,0}=E^{0}_{2,2,0}=E^{0}_{0,0,0}=
	-4J\left( h^0_{12}+h^0_{34} \right). 
\]
The triplet level is determined by eighteen eigenstates, where
$\lvert 1,1,1,m \rangle$ and $\lvert 2,2,1,m \rangle$, with $m=0,\pm1$, are 
characterized by the energy
\[
E^{m}_{2,2,1}=E^{m}_{1,1,1}=-3J\left( h^1_{12}+h^1_{34} \right)-
g\mu_{\mathrm{B}}Bm.
\]
Further, the respective triplets $\lvert 1,2,1,m \rangle$ and 
$\lvert 2,1,1,m \rangle$ energies are
\[
\begin{array}{c}
	E^{m}_{1,2,1}=-3J\left( h^1_{12}+h^1_{34} \right)-
	2J\left( h^1_{12}-h^1_{34} \right)-g\mu_{\mathrm{B}}Bm, \\ [0.2cm]
	E^{m}_{2,1,1}=-3J\left( h^1_{12}+h^1_{34} \right)+
	2J\left( h^1_{12}-h^1_{34} \right)-g\mu_{\mathrm{B}}Bm.
\end{array}
\] 
With $\lvert 0,1,1,m \rangle$ the Ni1-Ni2
coupled spin pair form a singlet. 
The set of constants $c^n_{12}$ in \eqref{eq:a00_ij} then can 
be determined in the limit of
zero external magnetic field, corresponding to $h^1_{12}=1$.
According to the estimations for the low-lying magnetic excitations
\cite{georgiev_epjb_2019}, found to be consistent
with inelastic neutron scattering experiments
\cite{nehrkorn_inelastic_2010,furrer_magnetic_PRB_2010}, we
have
$c^1_{12}=1.1923$ and $c^2_{12}=1.1153$, see e.g. Table \ref{tab:Param}. Therefore, with $m=m_{34}$ and $n=1,2$, 
for the positive sign in \eqref{eq:as0_ij} we get
\begin{equation}\label{eq:E^m_011}
	E^m_{0,1,1} \! \left( c^n_{12}\right) =-4Jc^n_{12}h^1_{12}-2Jh^1_{34}-
	g\mu_{\mathrm{B}}Bm
\end{equation}
and for the negative sign,
\begin{equation}\label{eq:E^0_011}
	E^{0}_{0,1,1} \! \left( c^n_{12}\right)=-4Jc^n_{12}h^1_{12}+2Jh^1_{34}.
\end{equation}
Thus, for each $m\ne0$ we get two energies since $n=1,2$ and for $m=0$ we have four.
On the other hand, when the Ni3-Ni4 spin pair is in the singlet state, \textit{i.e.} the triplet is 
$\lvert 1,0,1,m \rangle$, then $m=m_{12}$ and the coefficient 
$a^{1,0,0}_{34}=h^1_{34}c^n_{34}$. 
The analysis of Ni$_4$Mo$_{12}$ magnetic spectrum
was performed within the approximation $n=1$, where the
spectroscopic parameter $c^1_{34}=c_{34}=1.0384$. 
Consequently, accounting for the 
sign in \eqref{eq:as0_ij} we get
\[
E^{m}_{1,0,1}=-4Jc^{}_{34}h^1_{34}-2Jh^1_{12}-g\mu_{\mathrm{B}}Bm
\]
and 
\[
E^{0}_{1,0,1}=-4Jc^{}_{34}h^1_{34}+2Jh^1_{12}.
\]
The graphical representation of the energy sequence for $B=0$ is shown on 
Fig. \ref{fig:EnSpectrum}.

The quintet level is represented by thirty eigenstates and for $\lvert 1,1,2,m \rangle$ 
and $\lvert 2,2,2,m \rangle$, where $m=0,\pm1,\pm2$ we obtain
\[
E^{m}_{1,1,2}=E^{}_{2,2,2}=-J\left( h^2_{12}+h^2_{34} \right)
-g\mu_{\mathrm{B}}Bm. 
\]
When the cluster exhibits the structure of a local triplet and quintet spin bonds with
eigenstates $\lvert 1,2,2,m \rangle$ and $\lvert 2,1,2,m \rangle$ one obtains the 
following eigenvalues
\[
\begin{array}{c}
	E^{m}_{1,2,2}=-J\left( h^2_{12}+h^2_{34} \right)-
	2J\left( h^2_{12}-h^2_{34} \right)-g\mu_{\mathrm{B}}Bm, \\ [0.2cm]
	E^{m}_{2,1,2}=-J\left( h^2_{12}+h^2_{34} \right)+
	2J\left( h^2_{12}-h^2_{34} \right)-g\mu_{\mathrm{B}}Bm.
\end{array}
\]
Similarly to \eqref{eq:E^m_011} and \eqref{eq:E^0_011} when the spins of Ni1-Ni2
ions are paired in a singlet $\lvert 0,2,2,m \rangle$,
for the ``+'' sign in \eqref{eq:as0_ij} one gets
\[
E^{m}_{0,2,2}\! \left( c^n_{12}\right)=-4Jc^n_{12}h^2_{12}+2Jh^2_{34}-
g\mu_{\mathrm{B}}Bm
\]
and for the ``-'' sign
\[
E^{0}_{0,2,2}\! \left( c^n_{12}\right)=-4Jc^n_{12}h^2_{12}-2Jh^2_{34}.
\]
For the Ni3-Ni4 singlet the eigenvalues are
\[
E^{m}_{2,0,2}=-4Jc^{}_{34}h^2_{34}+2Jh^2_{12}-
g\mu_{\mathrm{B}}Bm
\]
and
\[
E^{0}_{2,0,2}=-4Jc^{}_{34}h^2_{34}-2Jh^2_{12}.
\]

The septet level consists of twenty
one eigenstates and in the presence of two
quintet spin bonds, $\lvert 2,2,3,m \rangle$, one gets the energy
\[
	E^{m}_{2,2,3}=2J\big( h^3_{12}+h^3_{34} 
	\big)-g\mu_{\mathrm{B}}Bm.
\]
The eigenvalues of the remaining fourteen eigenstates $\lvert 1,2,3,m \rangle$ and 
$\lvert 2,1,3,m \rangle$ are given by
\[
\begin{array}{c}
	E^{m}_{1,2,3}=2J\left( h^3_{12}+h^3_{34} \right)-
	2J\left( h^3_{12}-h^3_{34} \right) -g\mu_{\mathrm{B}}Bm, \\ [0.2cm]
	E^{m}_{2,1,3}=2J\left( h^3_{12}+h^3_{34} \right)+
	2J\left( h^3_{12}-h^3_{34} \right) -g\mu_{\mathrm{B}}Bm,
\end{array}
\]
respectively. The indirect splitting of the septet level related with the inequality 
$h^3_{12} \ne h^3_{34}$ is depicted on Fig. \ref{fig:EnSpectrum}. 

All nonet eigenstates $\lvert 2,2,4,m \rangle$, with total magnetic
quantum numbers 
$m=0,\pm1,\pm2,\pm3,\pm4$, correspond to the eigenvalue
\[
	E^{m}_{2,2,4}=6J\left( h^4_{12}+h^4_{34} \right)-g\mu_{\mathrm{B}}Bm.
\]

\subsection{Model parameters}\label{sec:magnetization}

The overall description of the magnetic properties of
Ni$_4$Mo$_{12}$ obtained with the aid of \eqref{eq:NickelHamilton}
requires fourteen parameters. Four field-independent parameters, i.e.
$J$, $c^1_{12}$, $c^2_{12}$ and $c_{34}$, are determined via
the neutron spectroscopic analysis, see e.g. Ref.
\cite{georgiev_epjb_2019}. The other ten
parameters, denoted commonly by ``$h$'', account for the indirect
influence of the externally applied magnetic field onto the correlation functions.
Their values can be determined from the magnetization and the magnetic susceptibility 
measurements.
Thus, one could expect that for small magnitudes of the external magnetic field
the proposed parameters will tend to one. 

We would like to point out that the parameters in 
\eqref{eq:AddHamiltonian} are intended to capture a particular degeneracy of the initial 
energy spectrum obtained within the framework of the applied variational method
briefly discussed in Sec. \ref{sec:spinmodel} and in more detail in
Ref. \cite{georgiev_epjb_2019}.
This degeneracy arises from the probability of observing different distributions 
of all electrons involved in the exchange processes. 
It is influenced by the presence of an externally applied magnetic field and it decreases exponentially with 
increasing temperature. Thus, the Hamiltonian in \eqref{eq:AddHamiltonian} is 
an effective model for studying low-temperature properties when $T\to0$ and 
any attempt of tuning the ``$h$'' parameters to high temperature measurements would 
lead to erroneous results.
%
Thereby the low-temperature and high-field magnetization measurements play a 
crucial role in extracting precise information on the values of the aforementioned parameters.

\begin{figure}[!h]
\includegraphics[scale=0.95]{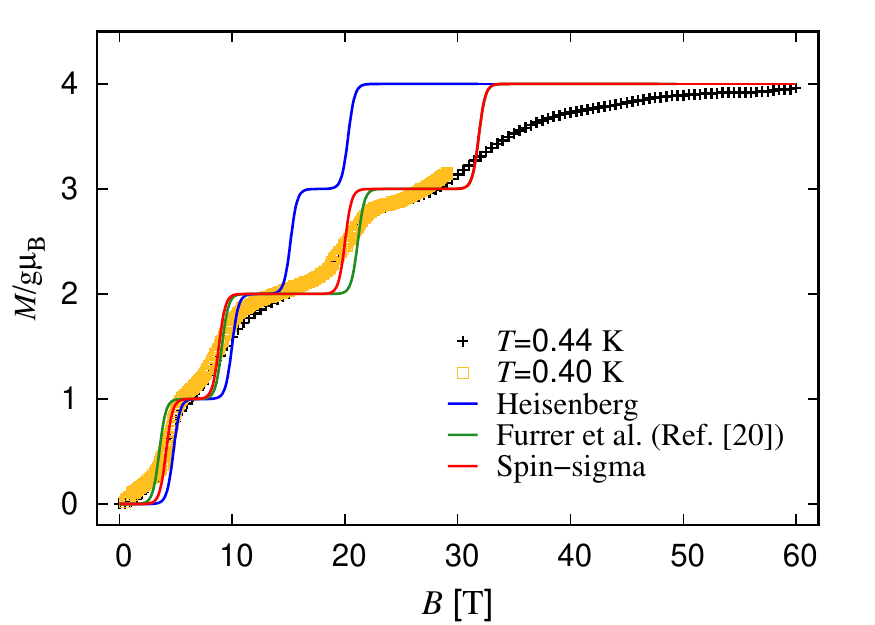}
\caption{Magnetization steps in the molecular magnet
Ni$_4$Mo$_{12}$.
The yellow and black symbols depict the experimental data from Ref.
\cite{schnack_observation_2006}. The solid blue and red lines
represent the calculated magnetization assuming the isotropic
Heisenberg and Hamiltonian \eqref{eq:AddHamiltonian}, respectively,
with $J=0.325$ meV.
The green line represents the magnetization curve obtained with the
aid of model \eqref{furrer2a} proposed in Ref.
\cite{furrer_magnetic_PRB_2010}.
All calculations are performed at $T=0.44$ K and $g=2.25$.
The fitted parameters of the Hamiltonian \eqref{eq:AddHamiltonian} are given in Table \ref{tab:Param}.
	\label{fig:MB}}
\end{figure}

\begin{figure}[!h]
\includegraphics[scale=0.95]{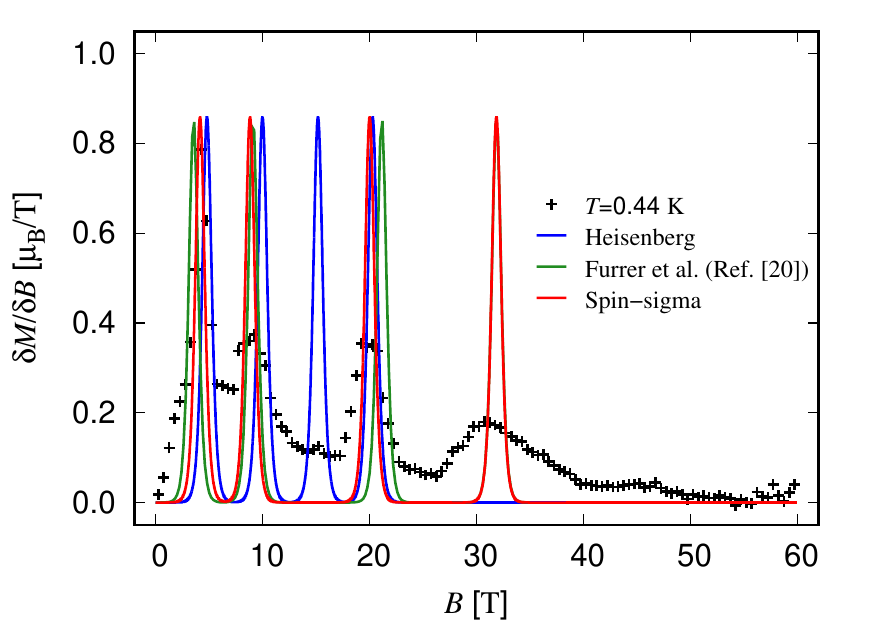}
\caption{Differential magnetization. The experimental data from Ref. 
\cite{schnack_observation_2006} are depicted with black symbols.
The calculated differential magnetization for the isotropic Heisenberg and 
Hamiltonian \eqref{eq:AddHamiltonian}, with $J=0.325$ meV,
are shown by blue and red lines, respectively.
The green line is associated to model \eqref{furrer2a} used in
Ref. \cite{furrer_magnetic_PRB_2010}.
All results are obtained at $T=0.44$ K and $g=2.25$.
The parameters entering into \eqref{eq:AddHamiltonian}
are provided in Table \ref{tab:Param}.
\label{fig:SB}}
\end{figure}

\begin{figure}[!h]
\includegraphics[scale=0.95]{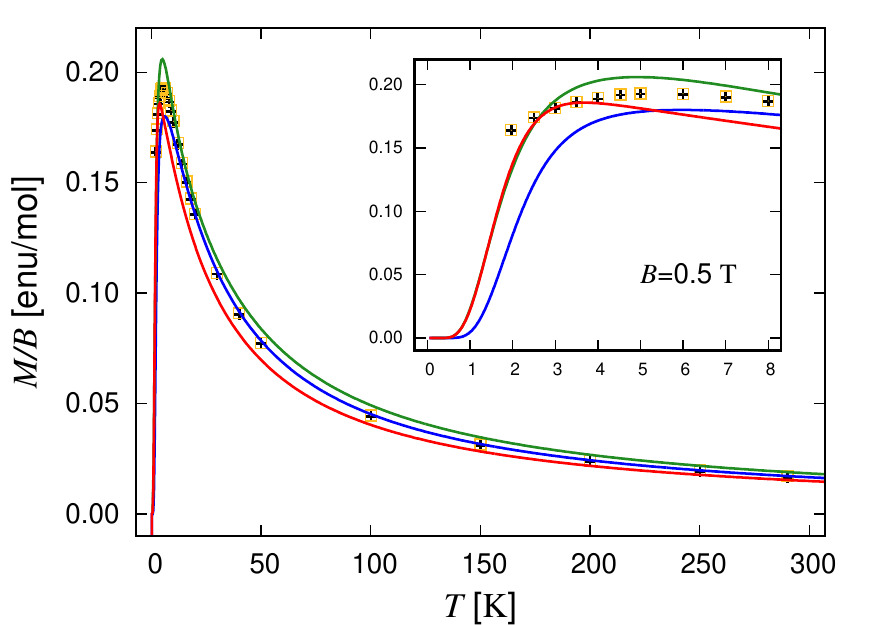}
\caption{Low-field susceptibility. The experimental points from Ref. 
\cite{schnack_observation_2006} are in yellow and black.
The blue and red lines represent the theoretical calculations, where the respective 
Hamiltonians are the isotropic Heisenberg and 
Hamiltonian \eqref{eq:AddHamiltonian} 
with $J=0.325$ meV.
The results obtained with the aid of model
\eqref{furrer2a} constructed
in Ref. \cite{furrer_magnetic_PRB_2010} are depicted by a green line.
All calculations are performed with respect to the values $B=0.5$ T and $g=2.25$.
The values of all field parameters for \eqref{eq:AddHamiltonian} are given in Table \ref{tab:Param}.
\label{fig:ST}}
\end{figure}

Using the Hamiltonian \eqref{eq:NickelHamilton} we have computed the magnetization, 
differential magnetization and low-field susceptibility shown 
on Figs. \ref{fig:MB}, \ref{fig:SB} and \ref{fig:ST}, respectively.
The values of the parameters $J$, $c^1_{12}$, $c^2_{12}$ and $c_{34}$
are determined according to
the neutron spectroscopy analysis reported in Ref. \cite{georgiev_epjb_2019} 
and are listed in Table \ref{tab:Param} along with the
values for the field parameters discussed in 
Sec. \ref{sec:energy}.

These values allow us to reproduce the experimentally observed
four magnetization steps found at $4.5$, $8.9$, $20.1$ and $32$ T of
Ref. \cite{schnack_observation_2006}.
The values of both parameters $h^0_{ij}$ and $h^1_{ij}$ indicate the low field dependence 
of Ni$_4$Mo$_{12}$ molecule.
On the other hand, the values of quintet and septet field parameters, corresponding to
the domain $10 \le B \le 30$, significantly differ form $h^0_{ij}$ and $h^1_{ij}$. 
Moreover, a pronounced jump is exhibited between $h^2_{ij}$ and $h^3_{ij}$. 
According to the used formalism such results
are a signal for relatively important electrons orbital contributions. 
As a result, independently of the Zeeman splitting the molecule exhibits shifting of the 
energy levels, see Fig. \ref{fig:EnSpectrum}.

\subsection{Comparison to model $H_{2a}$ of Ref. \cite{furrer_magnetic_PRB_2010}}
To describe the magnetic spectrum of the compound Ni$_4$Mo$_{12}$, in
Ref. \cite{furrer_magnetic_PRB_2010} several models based on the
conventional Heisenberg interaction among localized spins were proposed. One
of them tried to accommodate at the best the distorted crystalline structure of
the considered molecular magnet. Taking into account our convention,
the named model
involving anisotropic in space spin-spin interaction and an axial
single-ion anisotropy, reads
\begin{align}\label{furrer2a}
	H_{2a}=&2J\left(\hat{\mathbf{s}}_3\cdot\hat{\mathbf{s}}_1
	+\hat{\mathbf{s}}_3\cdot\hat{\mathbf{s}}_2
	+\hat{\mathbf{s}}_3\cdot\hat{\mathbf{s}}_4
	+\hat{\mathbf{s}}_1\cdot\hat{\mathbf{s}}_2\right) \nonumber \\
	& + 2J'\left(\hat{\mathbf{s}}_4\cdot\hat{\mathbf{s}}_1
	+\hat{\mathbf{s}}_4\cdot\hat{\mathbf{s}}_2\right) 
	+ D\sum_{i=1}^4\left(\mathrm{s}_i^z\right)^2.
\end{align}
Here, the parameters
$J=0.2517$ meV, $J'=0.5723$ meV and $D=0.2025$ meV
are determined by fitting to the relevant experimental data.


To compare our results, we computed the magnetization and the susceptibility with the aid of
Hamiltonian \eqref{furrer2a}.
The results for the associated physical quantities are depicted with a
green line on Figs.
\ref{fig:MB}, \ref{fig:SB} and \ref{fig:ST}. Notice that at weak external magnetic field,
Hamiltonians \eqref{eq:AddHamiltonian} and \eqref{furrer2a} lead to
slightly different results, while at strong fields ($B>25$ T) the corresponding curves are
indiscernible, see Figs. \ref{fig:MB} and \ref{fig:SB}. On Fig.
\ref{fig:ST}, we show the temperature dependence of the
low-field susceptibility at $B = 0.5$. We see clearly that at low temperatures our model \eqref{eq:AddHamiltonian}
fits better to experimental results than the Heisenberg Hamiltonian
\eqref{furrer2a} with single ion anisotropy.

\section{Conclusion}\label{sec:conclusion}

In order to study the magnetization and magnetic susceptibility of the
magnetic molecule
Ni$_4$Mo$_{12}$ we extended the introduced in Ref. \cite{georgiev_epjb_2019} 
post-Hartree-Fock method, based on the molecular orbital theory, by
accounting for the orbital contributions of the electrons to the intrinsic 
molecular field and found all electrons' correlations
as a field dependent \footnote[1]{M. Georgiev and H. Chamati, under preparation}.
We calculated the effect of externally applied magnetic field on the electrons' 
correlations with the aid of the effective spin-sigma bilinear form
\eqref{eq:AddHamiltonian} via the field parameters described in Sec. \ref{sec:sigma}.
This allowed us to reproduce the behavior of the magnetization, differential magnetization and low-field 
susceptibility data reported in Ref. \cite{schnack_observation_2006}.
As it is shown on Figs. \ref{fig:MB}, \ref{fig:SB} and \ref{fig:ST} the obtained results 
are consistent with the available experimental data.

The values of all parameters entering the theory are presented in Table \ref{tab:Param}. 
The role of the ``$h$'' parameters is to detect any variations in the zero-field energy spectrum 
induced from the externally applied magnetic field.
Thus, as the magnitude of the external magnetic field increases, 
the energy sequence alters making the molecule more resistive to the applied external action.
Consequently, one needs to input more magnetic energy in order to magnetize 
the molecule, Fig. \ref{fig:MB}. 
This process is also visible from the susceptibility measurements, see Fig. \ref{fig:SB}.
The gaps become wider increasing the magnitude of the external magnetic field and hence
making the Ni tetramer less susceptible.
On both Figs. \ref{fig:MB} and \ref{fig:SB} the described effect is
pronounced in the interval 10 -- 30 T.
%
%

This process is also imprinted in the diagram shown on Fig.
\ref{fig:EnSpectrum}.
The triplet and quintet energy levels are very close to each other and
almost indistinguishable.
As a consequence the triplet step, i.e. after $5$ and $8$ T, is not 
as wide as the upper steps. On the other hand, the quintet and
septet energy levels are separated by a larger energy gap, which
explains the extent of the quintet step. The same principle defines the width
of septet plateau.

The approach used of the present study is applicable to any isolated nanomagnetic 
unit with a set of magnetic centers interacting by one or more 
complex intermediate bridges. The microscopic model in \eqref{eq:AddHamiltonian} 
rests on the multi-configuration 
self consistent field method based on the molecular orbital theory.
Therefore, it is aimed to capture all magnetic features originating
from the electrons' delocalization and their distribution along the 
exchange bridges of a finite sized magnetic complex.
It may be of great value since a nontrivial bridging structure favors a multitude of
electrons' distributions leading to a number of magnetic excitations that don't
result from spin-orbit coupling processes nor from the existence of conducting bands.
Furthermore, such bridging structure also favors strong field dependent electrons' 
correlations arising form the orbital contribution of each delocalized electron 
included in the exchange processes.
In that respect neither of the known conventional effective spin models 
possess an adequate parametrization scheme. Moreover, any attempt to 
tune the exchange parameters from any such model lead to an incomplete
energy spectrum making the former inconvenient and thus inadequate.
To conclude, for a nanomagnets with trivial bonding bridges and localized 
electrons the Hamiltonian in \eqref{eq:AddHamiltonian} naturally reduces 
to the Heisenberg model.


\begin{acknowledgments}
The authors are indebted to Prof. N.S. Tonchev, Prof. N. Ivanov
for very helpful discussions, and to Prof. J. Schnack for providing 
us with the experimental data used in Figs. \ref{fig:MB}, \ref{fig:SB} and \ref{fig:ST}.
This work was supported by the Bulgarian National Science Fund under
Grant No. DN08/18 and the National program ``Young scientists and
postdoctoral researchers'' approved by DCM 577, 17.08.2018.
\end{acknowledgments}


%

\end{document}